\documentclass[showpacs,twocolumn, reprint,amsmath,amssymb, aps, prl]{revtex4-1}
\usepackage{amsmath}
\usepackage{amssymb}
\usepackage{float}
\usepackage{graphicx}
\usepackage{cancel}
\usepackage{color}
\usepackage{soul}
\usepackage{epstopdf}
\usepackage{dcolumn}% Align table columns on decimal point
\usepackage{bm}% bold math
\begin{document}
\newcommand{\beq}{\begin{equation}}
\newcommand{\eeq}{\end{equation}}
\newcommand{\bea}{\begin{eqnarray}}
\newcommand{\eea}{\end{eqnarray}}
\newcommand{\be}{\begin{equation}}
\newcommand{\ee}{\end{equation}}
\newcommand{\ba}{\begin{eqnarray}}
\newcommand{\ea}{\end{eqnarray}}
\def\mb{\mathbf}
\newcommand{\de}[1]{\textcolor{red}{#1}}

\title{
Phonon-mediated Casimir interaction between finite mass impurities
 }

\author{Andrei I. Pavlov}
\affiliation{Institute for Theoretical Solid State Physics, Leibniz-Institut f\"ur Festk\"orper-
und Werkstoffforschung IFW-Dresden, D-01169 Dresden, Helmholtzstraße 20,
Germany}
\author{Jeroen van den Brink}
\affiliation{Institute for Theoretical Solid State Physics, Leibniz-Institut f\"ur Festk\"orper-
und Werkstoffforschung IFW-Dresden, D-01169 Dresden, Helmholtzstraße 20,
Germany}
\author{Dmitri V. Efremov}
\affiliation{Institute for Theoretical Solid State Physics, Leibniz-Institut f\"ur Festk\"orper-
und Werkstoffforschung IFW-Dresden, D-01169 Dresden, Helmholtzstraße 20,
Germany}
\begin{abstract}
The Casimir effect, a two-body interaction via vacuum fluctuations,  is a fundamental property of quantum systems. In solid state physics it emerges as a long-range interaction between two impurity atoms via virtual phonons.
In the classical limit for the impurity atoms in $D$ dimensions the interaction is known to follow the universal power-law $U(r)\sim r^{-D}$. However, for finite masses of the impurity atoms on a lattice, it was predicted to be 
$U(r)\sim r^{-2D-1}$ at large distances.
We examine how one power-law can change into another with increase of the impurity mass and in presence of an external potential. We provide the exact solution for the system in one-dimension.
 %and show that it can be described by a two-length  scaling behavior. 
 At large distances indeed $U(r)\sim r^{-3}$ for finite impurity masses, while for the infinite impurity masses or in an external potential it crosses over to $U(r)\sim r^{-1}$ . At short distances the Casimir interaction is not universal and depends on the impurity mass and the external potential. 
\end{abstract}

\maketitle

Casimir in his pioneering work [\onlinecite{Casimir1948}] has shown that a
change of the zero-point energy due to	a perturbation of the electromagnetic fluctuations by two neutral metallic plates,  leads to  observable forces between these plates.
In fact, this is only one example of the broad class of phenomena, which are based on the concept of perturbation of the long-range fluctuations, e.g. Goldstone modes in the media with broken symmetry. 
Nowadays this effect, named after Casimir, 
can be encountered in various fields of physics, chemistry and biology \cite{Sparnaay1958,Polder1948,lifsh,dzyal,rev1,rev2,rev3}. For instance, in high energy physics the existence of Casimir effect sets natural  constraints on the Yukawa forces, appearing  due to the exchange of light elementary particles  and/or extra- dimensional physics \cite{part1}.  In cosmology the Casimir effect  helps to interpret the cosmological constant for a scalar field \cite{cosm1,cosm2,cosm3}. In chemistry, in particular, it is used to explain the interactions of molecules \cite{chem2,chem3}. In biology the Casimir interaction is for instance found to be responsible for organization of the bilayer structure of cell membranes \cite{biol1}.

    In condensed matter physics the effect of the Casimir interaction is extensively discussed with respect to interaction of conducting surfaces \cite{cond1}, graphene and conducting plates \cite{cond2}, mesoscopic particles in a critical fluid through critical fluctuations \cite{cond3} and ultracold atomic gases \cite{Waechter2007,Recati2005,Fuchs2007}.
    In the latter case it is possible to study the Casimir interaction in ultraclean bosonic or fermionic gases on tunable lattices with tunable spatial dimensionality and interaction strength. In this context one dimensional setups attract the most attention since the fluctuations are the strongest in 1D.

Precisely this situation was considered in [\onlinecite{Waechter2007,Recati2005,Fuchs2007}].  The authors studied the interaction between two static impurities due to  perturbation of phonon spectra in a Luttinger liquid. 
Since the mechanism is similar to the one proposed by Casimir, we hereafter denote it as the Casimir interaction.
The examination of the energy of zero-point motion of the Luttinger liquid in the presence of two impurities yielded the Casimir interaction  $U(r) \sim -1/r$. 
This dependence can be easily understood considering the zero-point energy of phonons in a potential well formed by two static impurities. The direct calculation leads to the following expression for the Casimir interaction \cite{Volovik,Zee}: 
\begin{equation} 
\label{casinf} U(r)=-\frac{c\pi}{24r}, 
\end{equation}
(here and below we use $\hbar =1$).

   At the same time, for two dynamical impurities which can move inside the medium, Schecter and Kamenev in [\onlinecite{kam}] proposed an essentially different $r$-dependence,   $U(r)= - mc^2 \frac{\Gamma_1 \Gamma_2}{32 \pi} \frac{\xi^3}{r^3}$, where $m$ is the mass of particles in the fluid, $c$ is the sound velocity, $\xi = 1/mc$ and the dimensionless parameters $\Gamma_{1,2}$ are impurity-phonon scattering amplitude. %The new power law was derived in the second order of the perturbation theory and higher order terms were summed up to the scattering amplitudes.   
   How the power law for dynamic impurities transforms to another for the static impurities is an open question. 
   
%   To address this question it is reasonable to investigate a simple model in which 
%   one can tune continuously dynamic impurities to static and check the evolution of the Casimir interaction.  For this reason  we study a harmonic crystal lattice with embedded two impurity neutral atoms.
   To address this question we investigate a model of a harmonic crystal lattice with embedded two impurity neutral atoms. It is arguably the simplest model in which 
   one can tune impurities continuously from dynamic to static and keep track of the evolution of the Casimir interaction. 
    In this model the Casimir interaction emerges naturally  
   between two impurity atoms as soon as their mass is different from the masses of the lattice atoms or an external potential is applied.   
%     It turns out that this model possesses an exact solution. 
%    Based on this solution we demonstrate how one power law transfers to another.
    %
%   Using the exact diagonalization method and the perturbation theory 
   We find that the Casimir interaction has different asymptotics 
   in these two cases. In the former one, for any finite mass of the impurity atoms 
   the Casimir interaction tends to the $1/r^3$-law at large distances in agreement with [\onlinecite{kam}].
   At the same time, in the limit of infinity mass the long range asymptotic tends to $1/r$ in agreement with [\onlinecite{Waechter2007,Recati2005,Fuchs2007}].
   In the case of an external potential the asymptotics is always $1/r$.
    
Our letter is organized as follows. 
%We start with the general consideration of possible interactions of phonons with impurities in the lattice model. 
Firstly we consider two neutral impurity atoms embedded in a harmonic crystal lattice. Using the exact diagonalization method we show that the power law at short distances strongly deviates from $1/r^3$ and the characteristic distance of the crossover to  $1/r^3$-law depends on the masses of the impurity atoms. Then we provide the exact solution for the model and formulate the continuum model. In the second part of the article formulate and exactly solve the model of two impurity atoms in an external harmonic potential. We show that the model is nonperturbative and has $1/r$ asymptotic behavior. Finally, we provide a discussion of the obtained results and conclusion.

\begin{figure}[t]
\centerline{ \includegraphics [width=1\linewidth] {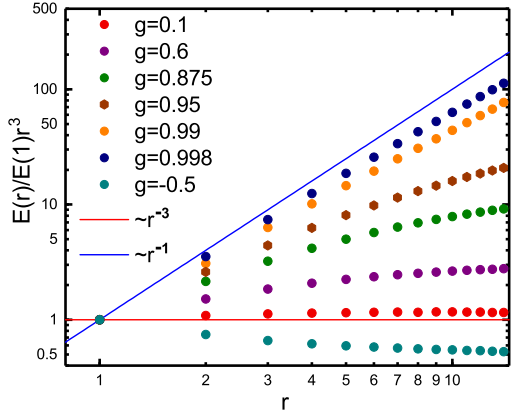} }
\caption{Normalized Casimir interaction $U_{\rm{eff}}(r)$ calculated for a chain of 200 atoms with two impurity atoms with various masses:
Red dots - $g=0.1$ $(M/m=1.1)$, purple - $g=0.6$ $(M/m=2.5)$, green - $g=0.875$ $(M/m=8)$, brown - $g=0.95$ $(M/m=20)$, orange - $g=0.99$ $(M/m=100)$, blue - $g=0.998$ $(M/m=500)$, turquoise - $g=-0.5$ $(M/m=0.5)$. The red line shows $1/r^3$ law, the blue line - $1/r$.}
\label{fig.1}
\end{figure}

\paragraph*{The model} --- 
We analyze an ideal harmonic cubic lattice described by
$
\nonumber H_0=\sum_i\frac{p_i^2}{2m}+ \frac{m\omega^2_0}{2}\sum_{<i,j>}(u_i-u_j)^2,
$
with two embedded   impurity atoms, which have mass or external potential different from the mass/potential of the atoms of the lattice. 
%\begin{equation}
% \label{eq.hamiltonian0}
%\end{equation}
Here $p_i$ and $u_i$ are the momentum and coordinate operators, $m$ is the mass of the atoms of the cubic lattice and $m\omega_0^2$ is the interaction potential.  

The Bogoliubov transformation brings $H_0$ to the Hamiltonian of noninteracting phonons:
\begin{equation}
\label{zeropoint} H_0 = \sum_{\mb{k}}  \omega_{\mb{k}}\left(b_{\mb{k}}^\dagger b_{\mb{k}} + \frac{1}{2}\right)
 \end{equation}
with the phonon spectrum:
$ \omega_{\mathbf{k}} = \omega_0 \sqrt{Z(1 - \gamma_\mb{k}) }$. 
  Here $\gamma_\mb{k} = \frac{1}{Z}\sum_{\Delta} e^{i \mb{k \Delta}}$ with summation over the nearest neighbours and $Z$ the number of the nearest neighbours. In one-dimensional case it reduces to:
$ \omega_k = 2 \omega_0\left|\sin (ka/2)\right|$,
 where $a$ is the lattice constant. In the low energy limit  $
 \omega_k = c|k|
 $ with the phonon velocity  $c = \omega_0 a$. Further for simplicity we put $a =1$.

 \paragraph{Two impurity atoms having different masses} --- Firstly, we consider two impurity atoms with masses $M$ located at the sites $a$ and $b$. The resulting Hamiltonian of the system is $H= H_0+V$  with the perturbation term of the kinetic energy: 
 \begin{equation}
 \mbox{~} V = -\frac{g}{2 m} (p^2_a + p^2_b ) 
 \label{eq.hamiltonianV}
 \end{equation}
 where the effective coupling constant $g=\left (1-m/M \right)$. 
 \paragraph*{Exact diagonalization} ---
The implication of two impurity atoms with masses $M$ breaks the translational invariance and $H$ can not be reduced to the Hamiltonian of free phonons. However, one can find the Casimir interaction, i.e  the dependence of the total energy of zero point motion  $E = \frac{1}{2}\sum_{ka}  \tilde\omega_{k}$ of the all atoms of the lattice on the distance between the impurity atoms. The result of exact diagonalization for 200 atom chain for various masses of impurity atoms is shown in Fig. \ref{fig.1}. To check the finite size effect we checked a twice large chain and found no difference. 

In general, the energy does not always fall down as $1/r^{3}$ as it was proposed in [\onlinecite{kam}] for Casimir effect in 1D. Moreover the interaction is not universal and depends on the mass of the impurity atoms (Fig. \ref{fig.1}). One can note that the normalized Casimir interaction for masses larger than $m$ is in the range $1/r^3<E<1/r$ for $r>1$, for light impurities ($M<m$) it is $E<1/r^3$. For impurity masses close to $m$, the Casimir interaction tends to $1/r^3$ law and in the limit $M\rightarrow\infty$  (static impurities) one observes the $1/r$ law. 
 
%\de{[I will Expand the discussion]}

%\begin{figure*}[t]
%\centerline{\includegraphics [width=0.5\linewidth] {fig2} \includegraphics [width=0.5\linewidth] {massesm}}
%\caption{a) The interaction $U(r)$ for $M/m=2$ obtained by exact diagonalization (blue points) and  in the second order of %perturbation theory (red points).
% Red line is the fitting $0.0025/r^3$. Blue line - function $0.0085/r^3$. b) Convergence of different space dependencies towards $\sim 1/r^3$. Purple dots - $m_{imp}=0.5m$, red dots - $m_{imp}=1.5m$, green dots - $m_{imp}=2m$, gray - $m_{imp}=2.5m$, brown - $m_{imp}=3m$, black - $m_{imp}=3.5m$.}
%\label{fig.2nd.order.vs.exact}
%\end{figure*}

\paragraph*{Perturbation theory}---
To find the reason of this drastic deviation of the distance dependence of the Casimir interaction from the $1/r^{3}$ law  we employ the perturbation theory. For the calculation we use the bosonic representation, in which Eq. (\ref{eq.hamiltonianV}) reads:
\begin{equation}\label{ham}  V= \sum_{\mb{q},\mb{q}'}(V^{(1)}_{\mb{q},\mb{q}'}b_\mb{q}^\dagger b_{\mb{q}'}+V^{(2)}_{\mb{q},\mb{q}'}\frac{b_\mb{q} b_{\mb{q}'}}{2}+h.c.). \end{equation}
Here the vertices are:
\bea  \nonumber V^{(1)}_{\mb{q},\mb{q}'}&=& -V^{(0)}_{\mb{q},\mb{q}'}
\cos\frac{(\mb{q}-\mb{q}')\mb{r}}{2}, \\
\nonumber V^{(2)}_{\mb{q},\mb{q}'}&= & V^{(0)}_{\mb{q},\mb{q}'}
\cos%\left( 
\frac{(\mb{q}+\mb{q}')\mb{r}}{2}
%\right),
%\label{eq.vertices.gamma1} 
\eea
with $\mathbf{r} = \mathbf{r}_a-\mathbf{r}_b $ and $V^{(0)}_{\mathbf{q},\mathbf{q}'} = g \sqrt{ \omega_q} \sqrt{ \omega_{q'}} $,
where   $\omega_q,\,\omega_{q'}$ are free phonon spectra given above. We choose $\mb{r}_a+\mb{r}_b=0$ for simplicity.
% Here we cancel m with the prefactor
%

The first order term of the perturbation theory is $r$-independent and therefore do not contribute to the Casimir interaction. The lowest order giving a contribution is the second order of the perturbation theory:
\begin{equation}
U_{\mbox{eff}}^{\!(2)}(r)\! =\! - %\int\limits_{-\infty}^{\infty}\!\!\frac{d\omega_n}{2\pi} 
2 T\sum_n
%\!\sum_{k,q}\! |\Gamma^{(0)}_{k,k+q}|^2\frac{\omega_k}{\omega_n^2\!+\!\omega^2_k} \frac{\omega_{k+q}}{\omega_n^2\!+\!\omega^2_{k+q}}\! \cos^2\!\!\left(\frac{qr}{2}\right).
\frac{|V^{(2)}_{k,k+q}|^2\omega_k \omega_{k+q}}{(\omega_n^2\!+\!\omega^2_k)(\omega_n^2\!+\!\omega^2_{k+q})}  .
\label{eq.second.order}
\end{equation}
%\de{[PLS check]\\}
Here $\omega_n=2\pi Tn$ is the Matsubara frequency.  

At large distances $r \gg 1$ the leading contribution comes from the small momenta. 
At zero temperature the integration Eq.(\ref{eq.second.order}) can be performed analytically for the linearized spectrum $\omega_k = ck$ with use of the substitution $T\sum_n\to\int d\omega_n/2\pi$. The result is the $1/r^3$-law:
\begin{equation} 
\label{en} 
U^{(2)}_{\rm{eff}}(r)= -\frac{g^2 \omega_0}{32\pi}\frac{1}{r^3}.
%\label{eq.second.order.of.perturbation.theory} 
\end{equation}
This dependence agrees with that previously found in [\onlinecite{kam}], but disagrees with the results of the exact diagonalization. 
\begin{figure}[t!]
	\centerline{
		\includegraphics [width=1\linewidth]{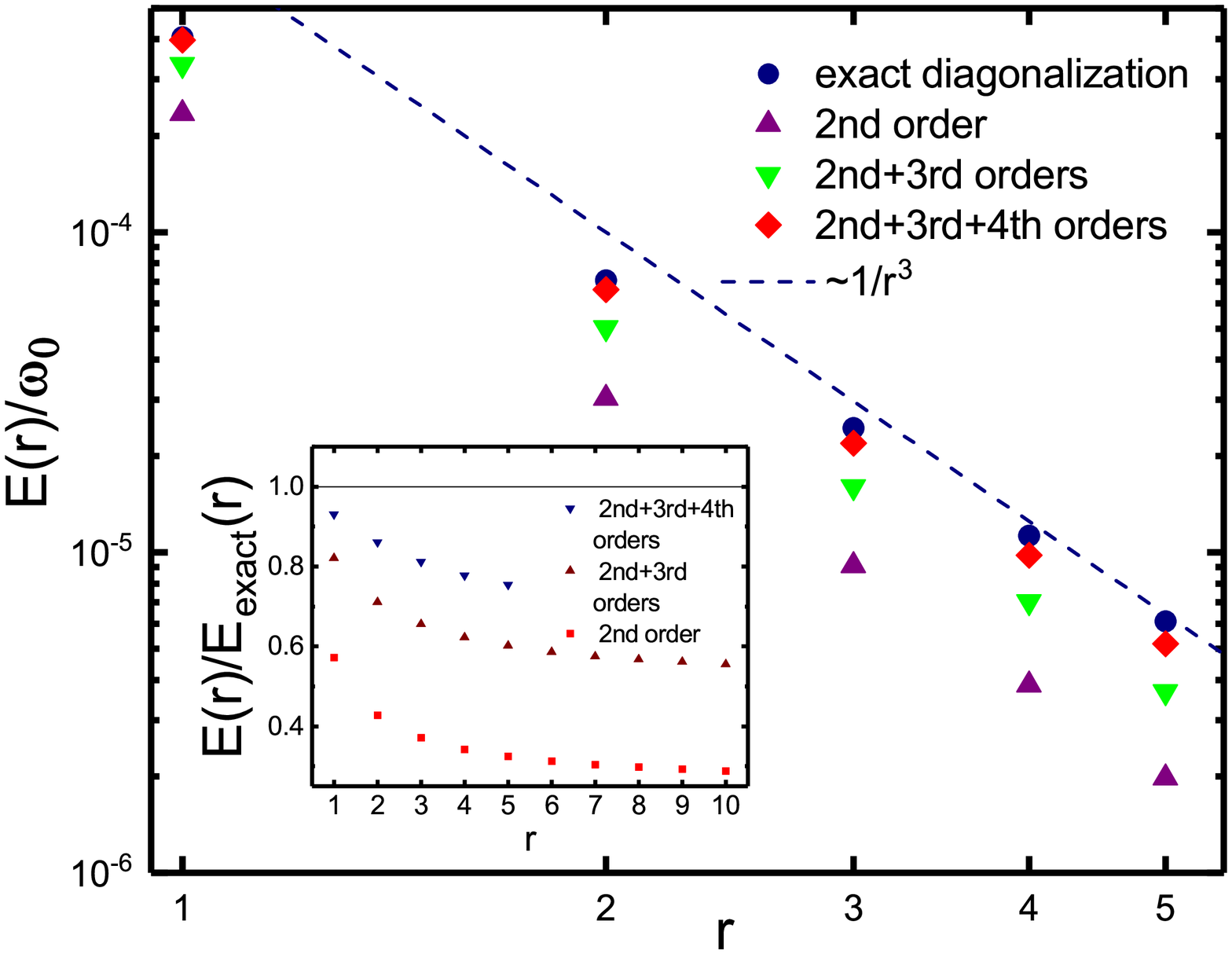} }
	\caption{Casimir interaction in the perturbation theory: gray dots - second order; brown dots - diagrams up to the third order; red dots - up to the forth order; blue dots - energies obtained by the exact diagonalization.
		Inset: Contribution of different orders of the perturbation theory to the total result.}
	\label{fig:totalord}
\end{figure}

\paragraph*{Higher order of perturbation theory} --- To understand the origin of the deviation from $1/r^3$ law, we explore higher order phonon processes, which correspond to multiple scattering of phonons on the impurities.
The result of the perturbation theory up to four-phonon processes for $g=0.5$ is presented in Fig. \ref{fig:totalord}. Here we keep only $r$-dependent terms. One immediately notes that the third and fourth orders of the perturbation theory significantly add to the Casimir interaction. Plotting the sum of these contributions up to the fourth order against the exact diagonalization reveals already a good match. 

\begin{figure}[t]
	\includegraphics [width=1\linewidth]{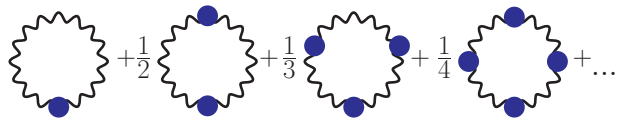}
	\caption{The diagrammatic representation of the thermodynamic potential.}
	\label{fig.thermodynamic.potential}
\end{figure}

The exact solution is given by the infinite sum of diagrams shown in Fig. \ref{fig.thermodynamic.potential}.
% where the boson spectrum is given by Eq.(\ref{eq.phonon.spectrum}) and the vertices by Eq.(\ref{eq.vertices.gamma1}). 
We can do this sum (see Supplementary material), and the obtained thermodynamic potential $\Phi(r)$ contains an  $r$-independent term, which is related to perturbation of the zero point motion by uncorrelated impurity atoms  ($r\to\infty$). 
%Therefore, to define the Casimir interaction one has to subtract this term.    
Defining $U_{\rm{eff}}(r)=\Phi(r)-\Phi(\infty)$ 
we arrive to the following
expression:
\begin{equation}  U_{\rm{eff}}(r)= \frac{1}{2} T\sum_n\ln\left[ 1-\left(\frac{gG(\omega_n,r)}{1-gG_0(\omega_n)}\right)^2\right], 
\label{eq.exact.solution}
\end{equation}
where $G(\omega_n,r)$ are the phononic Green functions in the coordinate space. 
%In the definition of $G(\omega_n,r)$ we transfer $k$-dependence from the vertex to the Green function (for details see \cite{Suppl}): 
Here we define the phononic field so, that $k$-dependence is transferred from the vertex to the Green function (for details see \cite{Suppl}):  
 \begin{eqnarray} \label{greenr}
 G(\omega_n,r) &=&
 \int^{\pi}_{-\pi}\frac{dk}{2\pi}\cos(kr)\frac{\omega^2_{k}}{\omega_n^2+\omega^2_{k}}
  \\ \nonumber
 &=&\!- \frac{\omega_0}{c} 
 f\left(\!\frac{|\omega_n|}{2\omega_0},r\!\right)\! ,
 \end{eqnarray}
 with

$ f(x,r) = \frac{x}{\sqrt{1+x^2}}(x + \sqrt{1+x^2})^{-2r}$, 
 and
\be
\label{greenapr} G_0(\omega_n)=2\int^{\pi}_{0}\frac{dk}{2\pi}\frac{\omega^2_{k}}{\omega_n^2+\omega^2_{k}}= 1-\frac{\omega_0}{c} f\left(\frac{|\omega_n|}{2\omega_0},0\right). 
\ee
One can note that the Green function for $r\gg 1$ decays exponentially fast $\sim e^{-2r\omega_n/\omega_0}$. 
It means that the main contribution to the Casimir interaction comes from the low energy acoustic phonons.   

\paragraph*{Continuum limit } --- The low energy Hamiltonian can be obtained from Eqs.(\ref{zeropoint},\,\ref{ham}) by  linearization of the spectrum for small momenta  $\omega_k = c|k|$. The corresponding Hamiltonian is:  
\begin{eqnarray} 
H &=& \sum_k c|k|  b_k^\dagger b_k 
\label{eq.hamiltonan.continous.kinetic}
\\&+& \!g c\! \sum_{k,k'} \!\!\sqrt{|k| |k'|}\!\cos\left[\frac{(k+k')r}{2}\right]\!\!\left(\!b_k^{\dagger}b_{k'}\!+\!\frac{b_k b_{k'} + b_k^{\dagger}b_{k'}^{\dagger} }{2}\!\right) \nonumber. 
\end{eqnarray}  
%
%The interaction term is similar to the \cite{AGD}
%
The only difference to the previous case is the change of the upper integration limit to infinity in Eqs.(\ref{greenr},\ref{greenapr}). Note that now the integral in Eq.(\ref{greenapr}) becomes divergent. The natural way of renormalization is the mapping on the lattice model. In this approach, at $T=0$ the Casimir energy reads:
\begin{equation} U_{\rm{eff}}(r)=T\sum_{n>0} %\int_{0}^{\infty}\frac{d\omega_n}{2\pi}
\ln\left[1-\left(\frac{\frac{g\omega_n}{2c}e^{-\frac{\omega r}{c}}}{1-g+\frac{g\omega_n }{2c}}\right)^2 \right] .
\label{eq.Casimir.linearized.T0}
\end{equation}
The direct comparison the results obtained with use of Eq.(\ref{eq.exact.solution}) and Eq.(\ref{eq.Casimir.linearized.T0}) for $r>1$ show excellent matching of the results \cite{Suppl}. From this expression it is clear that $U_{\rm{eff}}(r)$ decays exponentially fast at finite temperature for $r  \gg \omega_0/T$, i.e. thermo-fluctuations prevail on quantum fluctuations. The power law may emerge in some finite range.       
%The last integral can be estimated as:  
%\begin{equation} U_{\rm{eff}}(r)\simeq \frac{s c}{\pi r}\!\left(\!-\frac{1}{4s}+2(1+2s)\Gamma(0,4s)e^{4s}-1\!\right), \end{equation}
%where $s=\frac{1-g}{g}r$ and $\Gamma(0,4s)$ is the Gamma-function.
\begin{figure}[t]
	\includegraphics [width=1\linewidth]{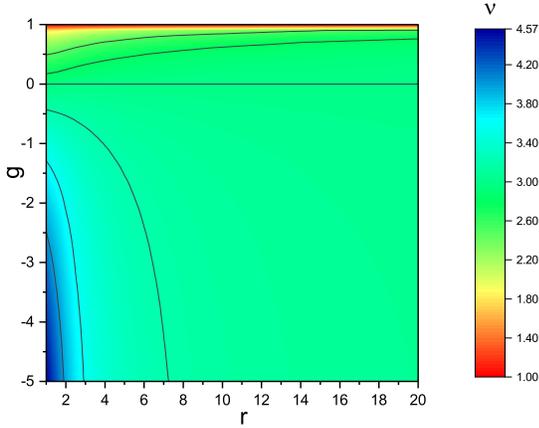}
	\caption{Logarithmic derivative $\nu=-\frac{d\ln U_{\rm{eff}}(r)}{d \ln r}$ as the function of $r$ and $g$ of the Casimir interaction between two impurity atoms having different masses.}
	\label{nu}
\end{figure}

To trace the dependence of the Casimir interaction on the coupling constant $g<1$ and distance $r$ at $T=0$ we introduce the logarithic derivative $\nu = -\frac{d \ln (E(r))}{d \ln (r)}$. For power law functions $1/r^\nu$ it gives the power $\nu$. The results are summarized in the Fig. \ref{nu}. The interval $0<g\leq 1$ describes the impurity masses $m<M\leq\infty$. The line $g=0$ is the singular line where $U_{\rm{eff}}=0$. And the interval $-\infty<g<0$ corresponds to $M<m$. 
One can see from the figure that although for small distances the Casimir interaction cannot be described by the  functions $1/r^\nu$, at large distances the dependence tends to $1/r^3$. 
The characteristic distance of the crossover to the $1/r^3$-law strongly depends on the masses of the impurity atoms. Finally, in the limit $g \rightarrow 1$ the Casimir interaction depends as $1/r$ from the distance between the impurity atoms and coincides with Eq. (\ref{casinf}).     

\paragraph{External potential} ---
Now we consider two atoms in an external harmonic potential which is defined by the following Hamiltonian:
\begin{equation} 
V=gm\omega_0^2(u_a^2+u_b^2), 
\label{eq.hamP}
\end{equation}
with the interaction constant $g \geq 0$.\\
%Vertices $\Gamma^{(1,2)}_{\mathbf{q},\mathbf{q}'}$ read as
%\begin{eqnarray}
%\Gamma^{(1)}_{\mb{q},\mb{q}'}&=& \Gamma^{(0)}_{\mb{q},\mb{q}'}e^{\textit{i}(\mb{q}-\mb{q}')\frac{\mb{R}}{2}}
%\cos[(\mb{q}+\mb{q}')\frac{\mb{r}}{2}],\\
%\Gamma^{(2)}_{\mb{q},\mb{q}'}&= & -\Gamma^{(0)}_{\mb{q},\mb{q}'}e^{\textit{i}(\mb{q}-\mb{q}')\frac{\mb{R}}{2}}
%\cos[(\mb{q}-\mb{q}')\frac{\mb{r}}{2}],
%\end{eqnarray} 
It leads to the new interaction term $V^{(0)}_{\mathbf{q},\mathbf{q}'}$:
\begin{equation}
\label{gammP} V^{(0)}_{\mb{q},\mb{q}'} = -\frac{g\omega_0^2}{\sqrt{ \omega_q} \sqrt{ \omega_{q'}} },
\end{equation} 

The bosonic Green functions are (see \cite{Suppl} for definition): 
\begin{eqnarray}
G(\omega_n,r)&=& \frac{\omega_0}{c}\frac{\omega_0^2}{\omega_n^2}f(|\omega_n|/2\omega_0, r) \label{eq.G.potential1}
\\
G_0(\omega_n)&=&\frac{\omega_0}{c}\frac{\omega_0^2}{\omega_n^2}f(|\omega_n|/2\omega_0, 0). 
\label{eq.G.potential2}
\end{eqnarray}

The direct calculation exhibits that all orders of the perturbation theory are divergent at the low energy limit \cite{Suppl}. But the summation of whole series of the diagrams Fig. \ref{fig.thermodynamic.potential}  leads to cancellation of the singularities and finite expression for the thermodynamic potential Eq.(\ref{eq.exact.solution}). The phononic Green functions are given by Eqs. (\ref{eq.G.potential1}-\ref{eq.G.potential2}).  

The correspondent continuous model is different from Eq. (\ref{eq.hamiltonan.continous.kinetic}) and is giving by: 
\begin{eqnarray} 
H &=& \sum_k c|k|  b_k^\dagger b_k 
\label{eq.hamiltonan.continous.potential}
\\&+& \!g\! \sum_{k,k'} \!\!\frac{\omega_0^2}{c\sqrt{|k k'|}}\!\cos\left[\frac{(k+k')r}{2}\right]\!\!\left(\!b_k^{\dagger}b_{k'}\!+\!\frac{b_k b_{k'} + b_k^{\dagger}b_{k'}^{\dagger} }{2}\!\right) \nonumber. 
\end{eqnarray}  

\begin{figure}[t!]
	\includegraphics [width=1\linewidth]{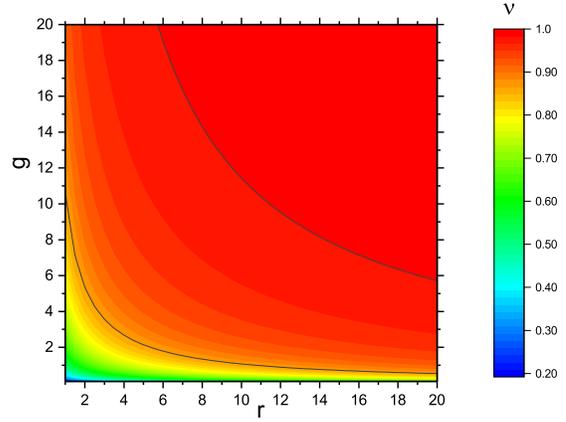}
	\caption{Logarithmic derivative $\nu=-\frac{d\ln U_{\rm{eff}}(r)}{d \ln r}$ as the function of $r$ and $g$ for the Casimir interaction between two masses in an external potential.}
	\label{densneg}
\end{figure}

The Casimir interaction has the form:
\begin{equation}
 \label{eq.thermodynamic.potential.energy} U_{\rm{eff}}(r)=
 %\int_0^{\infty}\frac{d\omega_n}{2\pi}
 T\sum_{n>0}\ln\left[ 1-\left(\frac{g\frac{\omega_0^2}{2c\omega_n}e^{-\frac{\omega_n}{c}r}}{1+\frac{g\omega_0^2}{2c\omega_n}}\right)^2\right]
\end{equation}

Similar expression was obtained in \cite{Recati2005}. To understand the scaling behavior at $T=0$ we plot the logarithmic derivative $\nu$ of the Casimir interaction $U_{\rm{eff}}$ given by Eq. (\ref{eq.thermodynamic.potential.energy}) as a function of  $r$ and $g$ in Fig. \ref{densneg}. For small values $gr$ the law is not universal, but  $U_{\rm{eff}}$ tends to $1/r$ as soon as $gr \ll 1$.    
The integral  Eq. (\ref{eq.thermodynamic.potential.energy}) in the limit $gr\gg 1$ matches the previously found expression for $M\to \infty$ Eq.(\ref{casinf}).

%can be evaluates in this limit:
%\begin{equation}
%U_{\rm{eff}}\approx-\frac{\omega_0^2 \pi}{24 c r}\left(1-\frac{2 c^2}{ gr\omega_0^2}+4(1+3\zeta(3)/\pi^2)\left(\frac{c^2}{gr\omega_0^2}\right)^2\right)\nonumber 
%\end{equation} 
%For the static impurities ($g\to\infty$) 
%it gives $U_{\rm{eff}} = \frac{\pi \omega_0^2}{24cr}$ in agreement with $M\to\infty$ in the previous case.

%
%\beq 
%\label{potEnA} U_{\rm{eff}}(r)\simeq \frac{g\omega_0^2 }{4\pi c}\left(1-\frac{g r \omega_0^2}{c^2}e^{g r \omega_0^2/c^2}\Gamma\left(0,\frac{g r \omega_0^2}{c^2}\right)\right) 
%\eeq
%%
%
%
%

\paragraph*{Discussion and conclusions} ---
The obtained long - range interaction can be observed experimentally  in ultra cold atomic gases as was shown in [\onlinecite{Recati2005}].  Since the competing Casimir-Polder interaction falls off much faster, namely as $1/r^6$, in the experimental setup of [\onlinecite{Moritz2005}] for the impurities at the distance of 1$\mu$m  the phonon induced Casimir interaction should dominate \cite{note1}.  

Summarizing, we have analyzed the evolution of the Casimir interaction between two impurity atoms embedded into an ideal 1D lattice at $T=0$. We have given the exact solution of the model and have studied the evolution of the Casimir interaction with change of the impurity atoms masses and the effect of an external potential. We have shown that multiboson processes change the scaling of the interaction decay with distance and the mass of the considered object plays an important role. As a consequence, the behavior at small distances differs from the power law at large. At large distances between two dynamic impurities the Casimir interaction is universal and obey $1/r^3$ law. For static impurities it tends to the $1/r$ law. 
%

%\footnote{ The minimal possible distance between the impurity atoms is 1 $\mu$m.    The authors of  [\onlinecite{Recati2005}] estimated the Casimir interaction between two static impurities for this setup as 1 kHz, which can be observed experimentally. For dynamic impurities one finds the interaction of the order of 1Hz. The Casimir-Polder interaction gives $10^{-6}$Hz for example for $^{40}$K atoms and $^{87}$Rb atoms.}

\paragraph*{Acknowledgments} --- We thank U. Nitzsche for technical assistance. D.V.E. and J.v.d.B would like to acknowledge the financial support provided by the German Research Foundation
(Deutsche Forschungsgemeinschaft) through the program DFG-Russia, BR4064/5-1. J.v.d.B is also supported by SFB
1143 of the Deutsche Forschungsgemeinschaft.

\bibliographystyle{apsrev}

\bibliography{casimir1}
\onecolumngrid
\appendix
%\begin{widetext}
\section{Supplemental Material}

\section{IMPURITY ATOMS WITH MASSES DIFFERENT FROM THE MASS OF THE
LATTICE ATOMS}

\subsection*{Perturbation theory up to the fourth order}

\begin{itemize} 
\item	Second order -- 
The second order perturbation term (Eq.(12) of the main text) integrated over frequency $\omega_n$ reads:
\beq \nonumber U_{\mbox{eff}}^{\!(2)}(r)\! =\! -\int\frac{dkdq}{(2\pi)^2}\frac{|V^{(0)}_{k,k+q}|^2}{\omega_k+\omega_{k+q}} \cos^2\!\left(\frac{qr}{2}\right). \eeq

\item  Third order reads:
\bea \nonumber U^{(3)}_{\mbox{eff}}(r)=-\frac{1}{4}\hbar\omega_0\int\frac{dq_1dq_2dq_3}{(2\pi)^3}\frac{|V^{(0)}_{q_1,q_2}V^{(0)}_{q_2,q_3}V^{(0)}_{q_1,q_3}|}{(\omega_1+\omega_2)(\omega_2+\omega_3)}\\
\nonumber *(\cos r(q_1+q_2)+\cos r(q_2+q_3)+\cos r(q_1-q_3)). \eea
\item  Forth order reads:

\bea \nonumber U^{(4,a)}_{\mbox{eff}}(r)&=&-\frac{1}{8}\hbar\omega_0\int\frac{dq_1dq_2dq_3dq_4}{(2\pi)^4}\frac{|V^{(0)}_{q_1,q_2}V^{(0)}_{q_3,q_4}V^{(0)}_{q_1,q_3}V^{(0)}_{q_2,q_4}|}{(\omega_{q_1}+\omega_{q_2})(\omega_{q_1}+\omega_{q_2}+\omega_{q_3}+\omega_{q_4})(\omega_{q_3}+\omega_{q_4})}[\cos r(q_1+q_2)\\
\nonumber &+&\cos r(q_3+q_4)+\cos r(q_1+q_3)+\cos r(q_2+q_4)+\cos r(q_1+q_4)\\
\nonumber &+&\cos r(q_2+q_3)+\cos r(q_1+q_2+q_3+q_4)] \eea
\bea \nonumber U^{(4,b)}_{\mbox{eff}}(r)&=&-\frac{1}{8}\hbar\omega_0\int\frac{dq_1dq_2dq_3dq_4}{(2\pi)^4}\frac{|V^{(0)}_{q_1,q_2}V^{(0)}_{q_1,q_3}V^{(0)}_{q_2,q_4}V^{(0)}_{q_3,q_4}|}{(\omega_{q_1}+\omega_{q_2})(\omega_{q_2}+\omega_{q_3})(\omega_{q_3}+\omega_{q_4})}[\cos r(q_1+q_2)\\
\nonumber &+&\cos r(q_3+q_4)+\cos r(q_2+q_3)+\cos r(q_1+q_4)+\cos r(q_1-q_3)\\
\nonumber &+&\cos r(q_2-q_4)+\cos r(q_1+q_2-q_3-q_4)] \eea
\bea \nonumber U^{(4,c)}_{\mbox{eff}}(r)&=&-\frac{1}{8}\hbar\omega_0\int\frac{dq_1dq_2dq_3dq_4}{(2\pi)^4}\frac{|V^{(0)}_{q_1,q_2}V^{(0)}_{q_2,q_3}V^{(0)}_{q_3,q_4}V^{(0)}_{q_1,q_4}|}{(\omega_{q_1}+\omega_{q_2})(\omega_{q_1}+\omega_{q_3})(\omega_{q_1}+\omega_{q_4})}[\cos r(q_1+q_2)\\
\nonumber &+&\cos r(q_1+q_3)+\cos r(q_1+q_4)+\cos r(q_2-q_3)+\cos r(q_2-q_4)\\
\nonumber &+&\cos r(q_3-q_4)+\cos r(q_1+q_2+q_3-q_4)] \eea

The forth order contains three nonvanishing at $T=0$ topologically nonequivalent diagrams (Fig.\ref{fig:forth}).\\

%\begin{figure}[t]
%	\centerline{
%		\includegraphics [width=1\linewidth]{dep} }
%	\caption{Exact result and lower orders of the perturbation theory for $g=0.5$.}
%	\label{fig:perturbationresults}
%\end{figure}

%All the corresponding values of $U^{(n)}_{\mbox{eff}}(r)$ can be calculated numerically (Fig. \ref{fig:perturbationresults}). But instead of that we use another approach that leads us to an analytical expression for the Casimir interaction in any order.\\
\begin{figure}[ht]
	\centerline{
		\includegraphics [width=0.5\linewidth]{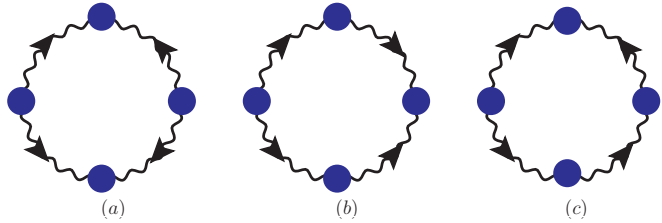} }
	\caption{Forth order diagrams}
	\label{fig:forth}
\end{figure}
\end{itemize}

\subsection*{Analytical solution} 
%Interaction between phonons and impurity atoms (see Eq. 8-11 from the main paper) contains momenta-dependence as a square root, that complicates the calculations. 
We use the definition of the phonon field $\varphi$ similar to used in [S1]:
\beq \nonumber \varphi(r,t)=\frac{1}{\sqrt{V}}\sum_q \sqrt{\omega_k}\left[ b_qe^{\textit{i}q r-\textit{i}\omega_q t}+b^+_qe^{-\textit{i}q r+\textit{i}\omega_q t} \right] \eeq
The phonon Green function in Matsubara formalism reads:
\beq \nonumber  D(q,\omega_n)=\omega_q\left( \frac{1}{\textit{i}\omega_n+\omega_q} +\frac{1}{-\textit{i}\omega_n+\omega_q}\right)\eeq 
%This result can be compared with the one given by logarithm expansion up to the third order in $g$ from the main paper. The procedure is done in the following way:
%\bea \nonumber U(r)&=&-\frac{1}{2}\int \frac{d\omega_n}{2\pi}\ln\left[ 1-\left( \frac{gG(\omega_n,r)}{1-gG_0(\omega_n)}\right)^2 \right]\simeq \frac{1}{2}\int\frac{d\omega_n}{2\pi}\frac{g^2G^2(\omega_n,r)}{1-2gG_0(\omega_n)+g^2G^2_0(\omega_n)}\\
%\nonumber &\simeq& \frac{1}{2}g^2\int\frac{d\omega_n}{2\pi}G^2(\omega_n,r)+\underline{g^3\int\frac{d\omega_n}{2\pi}G_0(\omega_n)G^2(\omega_n,r)}+... \eea

\begin{figure}[t!]
	\centerline{
		\includegraphics [width=0.5\linewidth]{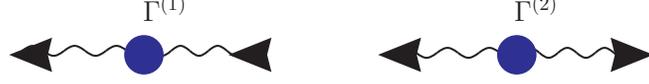} }
	\caption{Two type of vertices}
	\label{fig:vertices}
\end{figure}

Then the vertices of the phonon scattering on the impurities are  $\Gamma^{(1,2)}_{q,q'}=-g\cos[(q\pm q')\frac{r}{2}]$ (Fig. \ref{fig:vertices}).
% Since the vertices $\Gamma^{(2)}_{q,q'}$ and $\Gamma^{(2)\,*}_{q,q'}$ always appear in pairs, we can change their signs in front of them.\\
The basic block of any diagram is depicted in Fig. \ref{fig:green}:
\begin{figure}[t!]
	\centerline{
		\includegraphics [width=0.5\linewidth]{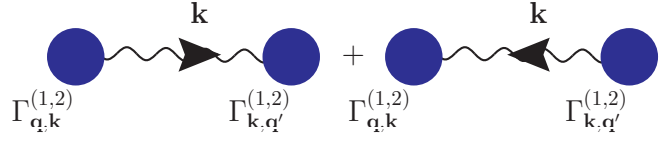} }
	\caption{Green function and two vertices}
	\label{fig:green}
\end{figure}
\bea \nonumber &&g^2\int_{-\pi}^{\pi}\frac{dk}{2\pi}\cos[(q\pm k)\frac{r}{2}]\cos[(k\pm q')\frac{r}{2}]\frac{2\omega_k^2}{\omega_n^2+\omega_k^2}\\
\nonumber &=&g^2\cos\frac{qr}{2}\cos\frac{q'r}{2}\int_{-\pi}^{\pi}\frac{dk}{2\pi}\cos^2\left(\frac{kr}{2}\right)\frac{2\omega_k^2}{\omega_n^2+\omega_k^2}+g^2\sin\frac{qr}{2}\sin\frac{q'r}{2}\int_{-\pi}^{\pi}\frac{dk}{2\pi}\sin^2\left(\frac{kr}{2}\right)\frac{2\omega_k^2}{\omega_n^2+\omega_k^2}\eea
%We will refer from now the two underlined expressions as Green functions:
It's worth to introduce Green functions in the coordinate space:
\bea \nonumber G_0(\omega_n)-G(r,\omega_n)=\int_{-\pi}^{\pi}\frac{dk}{2\pi}\cos^2\left(\frac{kr}{2}\right)\frac{2\omega_k^2}{\omega_n^2+\omega_k^2}\\
\nonumber G_0(\omega_n)+G(r,\omega_n)=\int_{-\pi}^{\pi}\frac{dk}{2\pi}\sin^2\left(\frac{kr}{2}\right)\frac{2\omega_k^2}{\omega_n^2+\omega_k^2} \eea
%\bea \nonumber -\left(G_0(\omega_n)-G(r,\omega_n)\right)=G_0(\omega_n)-G(r,\omega_n)\\
%\nonumber -\left(G_0(\omega_n)+G(r,\omega_n)\right)=G_0(\omega_n)+G(r,\omega_n) \eea
%Signs "$-$" in the left part cancel out with signs of vertices, giving us $gG_1(r, \omega_n)$ and $gG_2(r,\omega_n)$ products.\\
Then the loop of the $n$th order can be expressed in the compact form:
\beq \label{loops} U_{\mbox{eff}}^{\!(n)}=-\frac{1}{2}\frac{g^n}{n}\int_{-\infty}^{\infty}\frac{d\omega_n}{2\pi}\left((G_0(\omega_n)+G(r,\omega_n))^n +(G_0(\omega_n)-G(r,\omega_n))^n\right), \eeq
% $\frac{1}{2}$ multiplier comes, since with this definition of Green functions we count all topologically inequivalent diagrams twice. 
where:
\bea \nonumber
G(\omega_n,r)&=&\int^{\pi}_{-\pi}\frac{dk}{2\pi}\cos(kr)\frac{\omega^2_{k}}{\omega_n^2+\omega^2_{k}}=\int^{\pi}_{-\pi}\frac{dk}{2\pi}\cos(kr)\frac{4\omega_0^2\sin^2(\frac{k}{2})}{\omega_n^2+4\omega_0^2\sin^2(\frac{k}{2})}\\
\nonumber &=&\delta_{r,0}-\frac{\frac{|\omega_n|}{2c}}{\sqrt{1+(\frac{\omega_n}{2\omega_0})^2}} \left(\frac{\omega_n}{2\omega_0}+ \sqrt{1+\left(\frac{\omega_n}{2\omega_0}\right)^2}\right)^{-2r}\underset{\frac{\omega_n}{\omega_0}\ll 1}{\simeq} \delta_{r,0}-\frac{|\omega_n|}{2c}e^{-\frac{|\omega_n|}{\omega_0}r}\\
\nonumber G_0(\omega_n)&=&G(0,\omega_n)=\int^{\pi}_{-\pi}\frac{dk}{2\pi}\frac{\omega^2_{k}}{\omega_n^2+\omega^2_{k}}= 1-\frac{\frac{|\omega_n|}{2c}}{\sqrt{1+(\frac{\omega_n}{2\omega_0})^2}}\underset{\frac{\omega_n}{\omega_0}\ll 1}{\simeq} 1-\frac{|\omega_n|}{2c}. \eea

The thermodynamic potential at $T=0$ [1]:
\bea \nonumber \Phi_{total}(r)&=&-\int_0^{\infty}\frac{d\omega_n}{2\pi}\left[ \sum_{l=2}^{\infty}\frac{g^l}{l}\left((G_0(\omega_n)+G(r,\omega_n))^l +(G_0(\omega_n)-G(r,\omega_n))^l\right) \right]\\
\nonumber 
%&=&\int_0^{\infty}\frac{d\omega_n}{2\pi}\left[ \ln\left(1-gG_1(r,\omega_n) \right)+gG_1(r,\omega_n)+\ln\left(1-gG_2(r,\omega_n) \right)+gG_2(r,\omega_n) \right]\\
\nonumber
 &=&\int_0^{\infty}\frac{d\omega_n}{2\pi}\left[ \ln\left( 1-\frac{g^2G^2(r,\omega_n)}{(1-gG_0(\omega_n))^2} \right)+2\ln\left( 1-gG_0(\omega_n) \right)+2gG_0(\omega_n) \right] \eea
The effective Casimir energy goes to $0$ when $r\rightarrow\infty$ and should not contain a constant part:
\bea \nonumber U_{eff}(r)&=&\Phi_{total}(r)-\Phi_{total}(\infty)=\int_0^{\infty}\frac{d\omega_n}{2\pi}\left[ \ln\left( 1-\frac{g^2G^2(r,\omega_n)}{(1-gG_0(\omega_n))^2} \right)\right]\\
\label{energyexact} &=&\int_0^{\infty}\frac{d\omega_n}{2\pi}\ln\left[1-\frac{\left(\frac{g\omega_n}{2c}e^{-\frac{\omega_n}{c}r}\right)^2}{(1-g+\frac{g\omega_n}{2c})^2} \right],\eea
%=-\int_0^{\infty}\frac{d\omega}{2\pi}\ln\left[1-\frac{\left(\frac{\tilde{g}\omega}{2c}e^{-\frac{\omega}{c}r}\right)^2}{(1+\frac{\tilde{g}\omega}{2c})^2} \right]
\paragraph*{Continuum limit. }
The Green functions in the continuous limit can be obtained:
\bea \nonumber G(r,\omega_n)&\simeq& \int_{-\infty}^{\infty}\frac{dk}{2\pi}\cos kr\frac{k^2}{(\frac{\omega_n}{c})^2+k^2}=-\frac{|\omega_n|}{2c}e^{-\frac{|\omega_n|}{c}r}\\
\nonumber G_0(\omega_n)&=& 1-\int_{-\pi}^{\pi}\frac{dk}{2\pi}\frac{4\omega_0^2\sin^2\frac{k}{2}}{(\frac{\omega_n}{c})^2+4\omega_0^2\sin^2\frac{k}{2}}\simeq 1-\int_{-\infty}^{\infty}\frac{dk}{2\pi}\frac{(\frac{\omega_n}{c})^2}{(\frac{\omega_n}{c})^2+k^2}=1-\frac{|\omega_n|}{2c} \eea

%Let's write the expression for potential in the second order (only $r$-dependent part matters):
In the second order of the perturbation theory we restore the $1/r^3$ law:
\beq \nonumber U_{\mbox{eff}}^{(2)}(r)=
%-\frac{1}{2}\frac{g^2}{2}\int\frac{d\omega}{2\pi}(G_1^2\left(r,\omega_n))+G_2^2(r,\omega_n) \right)= 
-\frac{g^2}{2}\int_{-\infty}^{\infty}\frac{d\omega}{(2\pi)}\frac{\omega^2 e^{-2\frac{|\omega|r}{c}}}{4\omega_0^2}=-\frac{g^2\omega_0}{32\pi r^3}.
\eeq

\subsection*{Casimir Force}
The Casimir force reads:
%-\frac{dE_{eff}(r)}{dr}=-\int_0^{\infty}\frac{d\omega}{2\pi}\frac{\frac{g^2\omega^3}{2c^3}e^{-2\frac{\omega r}{c}}}{(1-g+g\frac{w}{2c})^2+(g\frac{\omega}{2c}e^{-\frac{\omega r}{c}})^2}=%
\bea \label{casimirforce} F(r)=-\frac{\partial U_{eff}}{\partial r}=-\int_0^{\infty}\frac{d\omega}{2\pi}\frac{\frac{\tilde{g}^2\omega^3}{2c^3}e^{-2\frac{\omega r}{c}}}{(1+\tilde{g}\frac{\omega}{2c})^2-(\tilde{g}\frac{\omega}{2c}e^{-\frac{\omega r}{c}})^2}, \eea
where we use a new constant $\tilde{g}=\frac{g}{1-g}$ for convenience. For heavy impurities ($\tilde{g}>0$), one can approximate Eq. (\ref{casimirforce}) omitting the exponentially small term from the denominator. It reads then as:
\beq \label{force} F(r)\simeq -\int_0^{\infty}\frac{d\omega}{2\pi}\frac{\frac{\tilde{g}^2\omega^3}{2c^3}e^{-2\frac{\omega r}{c}}}{(1+\tilde{g}\frac{w}{2c})^2}=\frac{c}{\tilde{g}^2\pi}\left(\frac{1}{4 x^2}-\frac{2}{x}-4+(12+16x)I(4x)\right), \eeq
where $x=r/\tilde{g}$. Here $I(x)=\int_0^{\infty}dt\frac{e^{-t}}{x+t}$. It can be expressed through the incomplete gamma function: $I(x)=e^x\Gamma[0,x]$, $\Gamma[\alpha,x]=\int_x^{\infty}t^{\alpha-1}e^{-t}dt$.

\begin{figure}[t]
\includegraphics [width=1\linewidth]{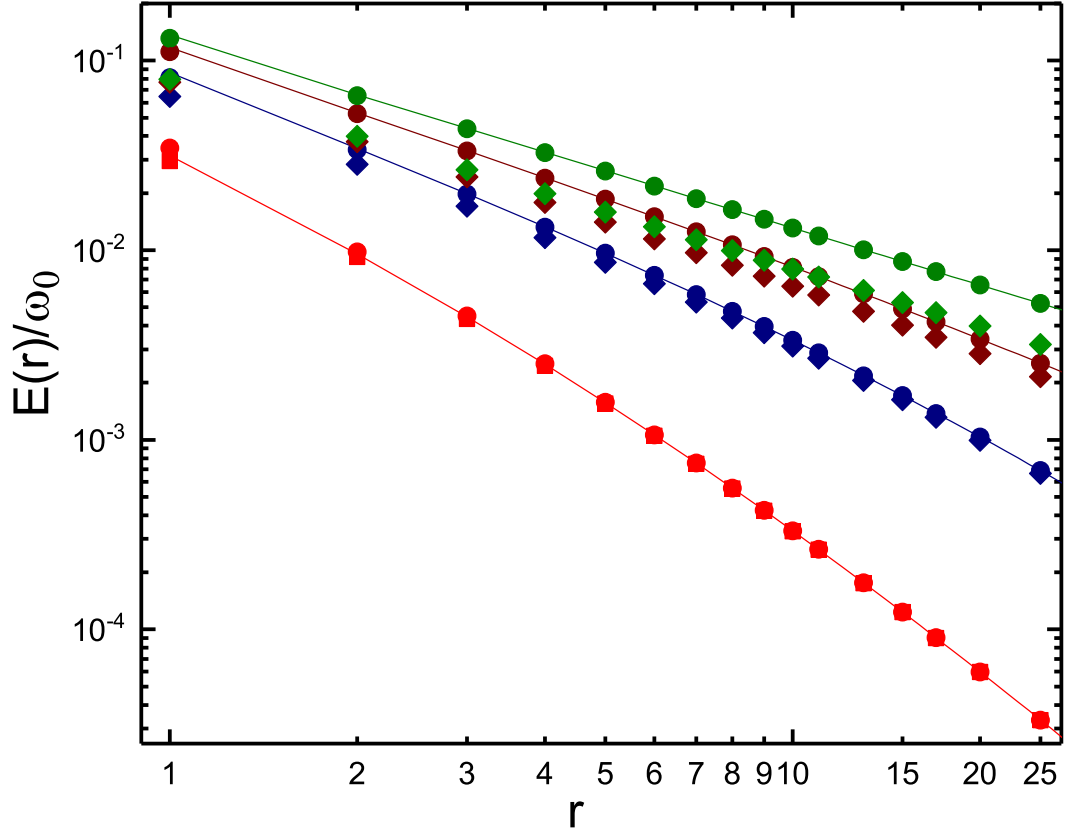}
\caption{Comparison of exact result, result for the linearized vertices and approximate analytical formula. Red color - $g=0.9$, blue - $g=0.99$, brown - $g=0.999$, green - $g=1$. Circles - exact result, lines - linearized vertices, diamonds - approximate formula.}
\label{cmp}
\end{figure}

\begin{figure}[t]
\centerline{
\includegraphics [width=1\linewidth]{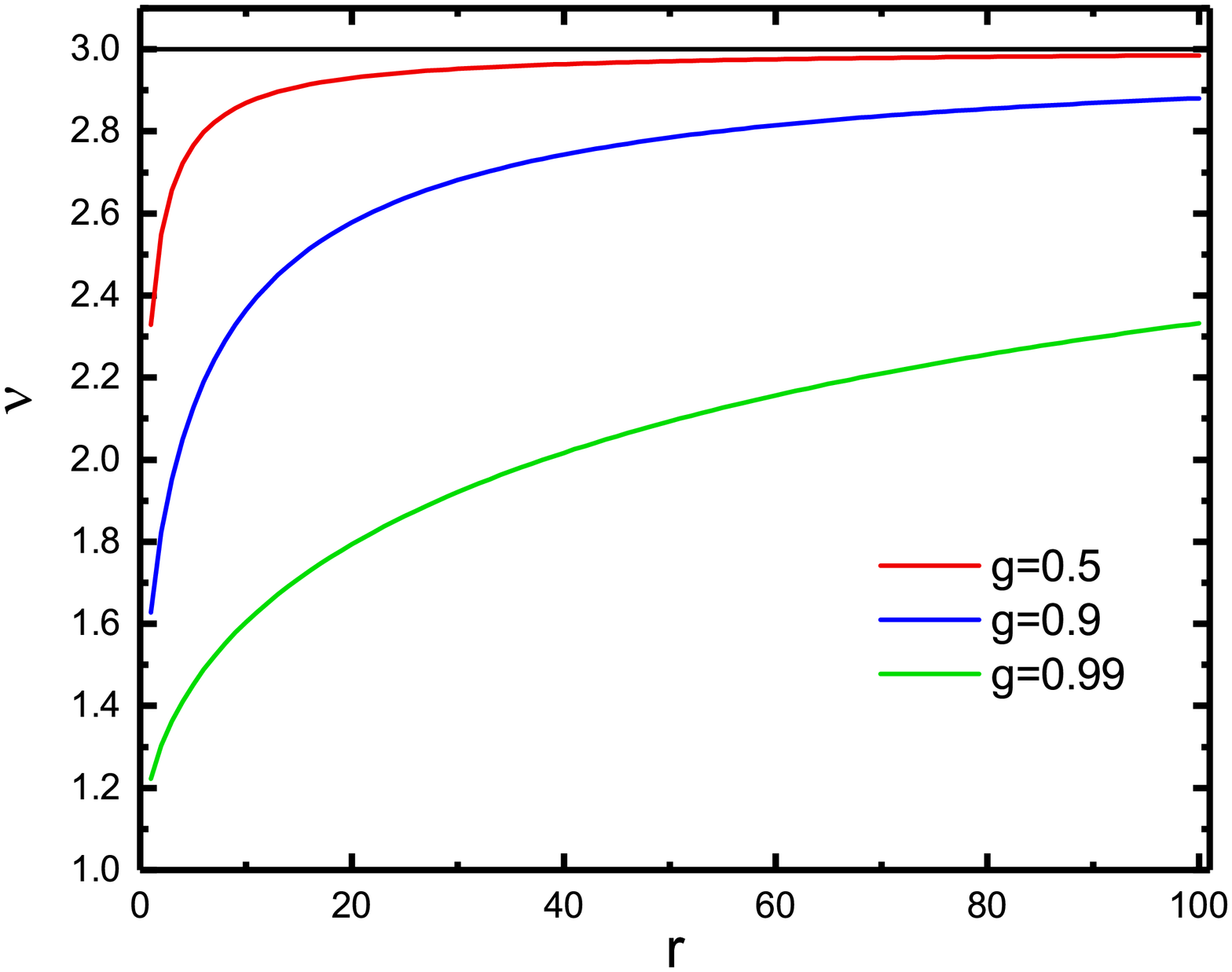} }
\caption{Dependence of the logarithmic derivative $\nu(r)$ on distance for various fixed $g$.}
\label{pw}
\end{figure}

%For $x\gg1$, it behaves asymptotically as
%\beq F(x)\underset{x\rightarrow \infty}{\simeq}\frac{c}{\tilde{g}^2\pi}(\frac{3}{32x^4}-\frac{3}{16x^5}+\frac{45}{128x^6}+...). \eeq 
%For $x\ll1$:
%\beq F(x)\underset{x\rightarrow 0}{\simeq}\frac{c}{\tilde{g}^2\pi}(\frac{1}{4x^2}-\frac{2}{x}-144\ln x-192x\ln x +...). \eeq
%Eq. (\ref{force}) can be integrated, giving us 
%\beq \nonumber U(x)\simeq-\frac{c}{\tilde{g}\pi}(-\frac{1}{4x}+2e^{4x}(1+2x)\Gamma[0,4x])+C \eeq
%Taking into account that $U(\infty)=0$, we find that $C=-\frac{1}{\pi}$, so the Casimir interaction energy can be written as
Integration over $r$ with condition $U_{\rm{eff}}(r\rightarrow\infty)=0$ gives:
\beq \label{energyapprox} U_{\rm{eff}}(x)\simeq \frac{c}{\tilde{g}\pi}\bigg(-\frac{1}{4x}+2(1+2x)I(4x)-1\bigg) \eeq
Expression (\ref{energyapprox}) works excellent for small masses, but for infinite masses it gives $\frac{1}{4\pi}$ numerical coefficient instead of $\frac{\pi}{24}$ provided by (\ref{energyexact}) and expected for the Casimir law (Fig. \ref{cmp}).
Asymptotically, Eq. (\ref{energyapprox}) for mass ratio $m/M\rightarrow 1 (g\rightarrow 0)$ is:
\beq \nonumber U_{\rm{eff}}(x)\underset{x\rightarrow \infty}{\simeq}\frac{c}{\tilde{g}}\bigg(-\frac{1}{32\pi x^3}+\frac{1}{16\pi x^4}-\frac{9}{128\pi x^5}+...\bigg) \eeq
%\beq \nonumber E(x)\underset{x\rightarrow 0}{\simeq}\frac{c}{\tilde{g}}(-\frac{1}{4\pi x}-\frac{2}{\pi}\ln x-\frac{12}{\pi}x\ln x) \eeq 
The space dependence for this expression is shown at FIG. \ref{pw} with three different values of $g$.

%\begin{figure}[H]
%\includegraphics [width=1\linewidth]{density2}
%\caption{Logarithmic derivative $\nu$ for various values of $r$ and $g$.}
%\label{interaction}
%\end{figure}

\section*{External potential} 
\begin{figure}[H]
\centerline{
\includegraphics [width=1\linewidth]{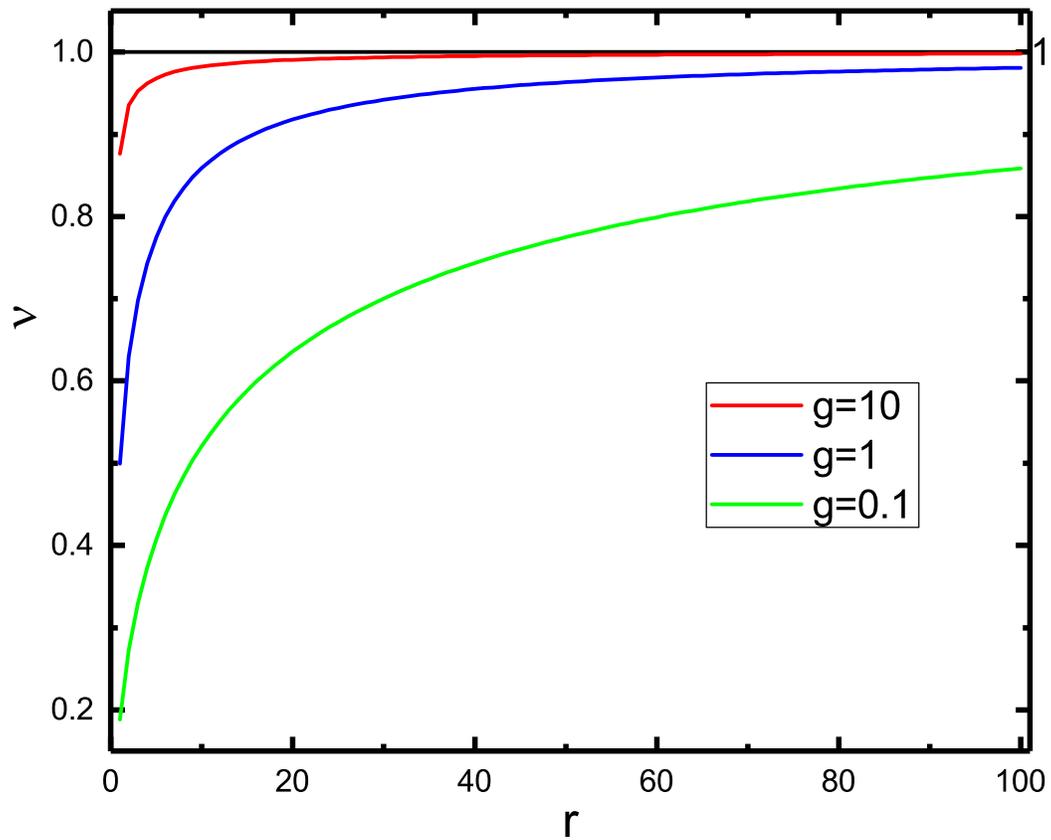} }
\caption{Dependence of the logarithmic derivative $\nu(r)$ on distance for various fixed $g$. Potential energy case.}
\label{fig:pw1}
\end{figure}

%It is also possible to consider another model. We suppose that introduction of a different kind of atoms on the lattice does not affect their kinetic energy, changing their potential energy instead. We still have the harmonical oscillator Hamiltonian (Eq. 3 in the main paper), but 
We consider two atoms in an external potential $V$ given by:
\beq \label{eq.hamP} V=gm\omega_0^2(u_a^2+u_b^2), \eeq
$g\in(0,\infty)$.\\
For the calculation we use the bosonic representation, in which the perturbation reads:
\begin{equation}\label{ham}  V= \sum_{\mb{q},\mb{q}'}(V^{(1)}_{\mb{q},\mb{q}'}b_\mb{q}^\dagger b_{\mb{q}'}+V^{(2)}_{\mb{q},\mb{q}'}\frac{b_\mb{q} b_{-\mb{q}'}}{2}+h.c.). \end{equation}
Here the vertices are:
\bea \nonumber V^{(1)}_{\mb{q},\mb{q}'}&=& V^{(0)}_{\mb{q},\mb{q}'}
\cos[(\mb{q}-\mb{q}')\frac{\mb{r}}{2}],\\
\nonumber V^{(2)}_{\mb{q},\mb{q}'}&= & -V^{(0)}_{\mb{q},\mb{q}'}
\cos[(\mb{q}+\mb{q}')\frac{\mb{r}}{2}],\eea
with $V^{(0)}_{\mathbf{q},\mathbf{q}'}$:
\beq \label{gammP} V^{(0)}_{\mb{q},\mb{q}'} = \frac{g\omega_0^2}{\sqrt{ \omega_q} \sqrt{ \omega_{q'}} }.\eeq
Now we define a free phonon field as 
\beq \nonumber \tilde{\varphi}(r,t)=\frac{1}{\sqrt{V}}\sum_q \omega_0\sqrt{\frac{1}{\omega_k}}\left[ b_qe^{\textit{i}q r-\textit{i}\omega_q t}+b^+_qe^{-\textit{i}q r+\textit{i}\omega_q t} \right] \eeq
The Green functions $G(\omega_n,r), G_0(\omega_n)$ take form:
\bea \nonumber
G(\omega_n,r)&=&\int^{\pi}_{-\pi}\frac{dk}{2\pi}\cos(kr)\frac{\omega^2_{0}}{\omega_n^2+\omega^2_{k}}=\int^{\pi}_{-\pi}\frac{dk}{2\pi}\cos(kr)\frac{\omega_0^2}{\omega_n^2+4\omega_0^2\sin^2(\frac{k}{2})}\\
\nonumber &=&\frac{\frac{\omega_0^2}{2|\omega_n|c}}{\sqrt{1+(\frac{\omega_n}{2\omega_0})^2}}\left(\frac{\omega_n}{2\omega_0}+\sqrt{1+\left(\frac{\omega_n}{2\omega_0}\right)^2}\right)^{-2r}\\
\nonumber G_0(\omega_n)&=&G(0,\omega_n)=\int^{\pi}_{-\pi}\frac{dk}{2\pi}\frac{\omega^2_{0}}{\omega_n^2+\omega^2_{k}}= \frac{\frac{\omega_0^2}{2|\omega_n|c}}{\sqrt{1+(\frac{\omega_n}{2\omega_0})^2}}. \eea 
%Using these expressions for (\ref{energyexact}), we can calculate energies and logarithmic derivatives $\nu$ for any positive $r$ and $g$ (for negative $g$ the problem becomes unstable). $\nu$ is presented at Fig. 5 in the main paper. 
To analyze the Casimir energy, we use the linearized spectrum. Green functions read as
\bea \nonumber G(\omega_n,r)&=&\frac{\omega_0^2}{2c|\omega_n|}e^{-\frac{|\omega_n|}{c}r},\\
\nonumber G_0(\omega_n)&=&\frac{\omega_0^2}{2c|\omega_n|}. \eea
It's worth to consider the second order term of the perturbation theory for the Casimir interaction:
\beq \nonumber U^{(2)}=-\frac{g^2}{2}\int_{-\infty}^{\infty}\frac{d\omega_n}{2\pi}\frac{\omega_0^4}{4c^2\omega_n^2}e^{-2\frac{|\omega_n|}{c}r} \eeq
This expression diverges at small frequencies. All other terms diverge as well.\\ 
But the whole sum of the perturbation theory series remains finite and gives us:
\beq \label{potEn} U_{eff}(r)=-\int_0^{\infty}\frac{d\omega_n}{2\pi}\ln\left[ 1-\left(\frac{g\frac{\omega_0^2}{2c\omega_n}e^{-\frac{\omega_n}{c}r}}{1+\frac{g\omega_0^2}{2c\omega_n}}\right)^2\right]
\eeq
The distance dependence for various given values of $g$ is shown at Fig. \ref{fig:pw1}.\\

At $g=\infty$, from Eq.(\ref{potEn}) follows $U_{eff}(r)=\frac{\omega_0^2\pi}{24cr}$. To investigate the finite $g$ case, we use the same approach as before, finding an approximate expression for the Casimir force and integrating it:
\bea \nonumber F(x)&=&\int_0^{\infty}\frac{d\omega_n}{2\pi}\frac{g^2\frac{\omega_0^4}{2c^3\omega_n}e^{-2\frac{\omega_n}{c}r}}{(1+\frac{g\omega_0^2}{2c\omega_n})^2-\frac{g^2\omega_0^4}{4c^2\omega_n^2}e^{-2\frac{\omega_n^2}{c}r}}\\
\label{forcepot} &\simeq& \frac{g^2\omega_0^4}{4\pi c^3}\left(-1+(x+1)I(x)\right), \eea
where $x=\frac{g r \omega_0^2}{c^2}$.
\beq \label{potEnA} E_{Cas}(r)\simeq \frac{g\omega_0^2 }{4\pi c}\left(1-xI(x)\right) \eeq
At Fig. \ref{densneg}, distribution of energy (\ref{potEnA}) in relation to $r$ and $g$ is shown. The $r$-dependence for various given values of $g$ is also depicted at Fig. \ref{fig:pw1}.\\ 
%At $g=\infty$, the energy turns into $\frac{\omega_0^2}{4\pi c r }$, while the exact expression (\ref{potEn}) gives $\frac{\omega_0^2\pi}{24c r}$. Unlike the previous case of kinetic energy change, (\ref{potEnA}) gives us a numerical prefactor which differs from (\ref{potEn}) for all values of $g$, though $r$ and $g$ dependencies are correct. It happens due to the fact that small $\omega$ region in (\ref{forcepot}) contributes sufficiently for the last term in the denominator. In the case of $gr\gg 1$, it is better to rewrite the expression (\ref{forcepot}) in the following way:
%\beq \nonumber F(r)=\frac{\omega_0^2}{c r^2}\int_o^{\infty}\frac{dx}{2\pi}\frac{2x e^{-2x}}{(1-e^{-2x})+\frac{4x c^2}{gr\omega_0^2}+(\frac{2x c^2}{gr\omega_0^2})^2} \eeq
For $gr\gg 1$, the expression for the Casimir force reads:
\beq \nonumber F(r)=\frac{\omega_0^2}{c r^2}(\frac{\pi}{24}-\frac{c^2 \pi}{6 gr\omega_0^2}+\frac{\pi^2+3\zeta(3)}{2\pi}(\frac{c^2}{gr\omega_0^2})^2+...)\nonumber 
\eeq

[S1] A. A. Abrikosov, L. P. Gorkov, I. Y. Dzyaloshinskii, \textit{Methods of Quantum Field Theory in Statistical Physics} (Dover Publications, New York, 1975).
%\end{widetext}

\end{document}